\documentclass[sigconf]{acmart}
\usepackage{multirow} 
\usepackage{algorithm}
\usepackage{algpseudocode} 
\AtBeginDocument{%
  }

\setcopyright{none}
\copyrightyear{2025}
\acmYear{2025}
\acmConference[FAccTRec'25]{FAccTRec Workshop on Responsible Recommendation}{September 26, 2025}{Prague, Czech Republic}
  
\begin{document}

\title{A Case Study of Balanced Query Recommendation on Wikipedia}

\author{Harshit Mishra}
\email{hamishra@syr.edu}

\affiliation{%
  \institution{Syracuse University}
 \city{Syracuse}
  \state{New York}
  \country{USA}
}

\author{Sucheta Soundarajan}
\email{susounda@syr.edu}
\affiliation{%
  \institution{Syracuse University}
  \city{Syracuse}
  \state{New York}
  \country{USA}
}


\begin{abstract}
  Modern IR systems are an extremely important tool for seeking information. In addition to search, such systems include a number of query reformulation methods, such as query expansion and query recommendations, to provide high quality results. However, results returned by such methods sometimes exhibit undesirable or wrongful bias with respect to protected categories such as gender or race.  Our earlier work considered the problem of \textit{balanced query recommendation}, where instead of re-ranking a list of results based on fairness measures, the goal was to suggest queries that are relevant to a user’s search query but exhibit less bias than the original query. 
  In this work, we present a case study of \texttt{BalancedQR} using an extension of \texttt{BalancedQR} that handles biases in multiple dimensions. It employs a Pareto front approach that finds balanced queries, optimizing for multiple objectives such as gender bias and regional bias, along with the relevance of returned results. We evaluate the extended version of \texttt{BalancedQR} on a Wikipedia dataset.  Our results 
  demonstrate the effectiveness of our extension to \texttt{BalancedQR} framework and highlight the significant impact of subtle query wording, linguistic choice on retrieval. 
\end{abstract}




\maketitle

\section{Introduction}
Online search engines and recommender systems are a vital part of the modern information retrieval ecosystem. However, as is well-known, the results returned by search engines may over- or under-represent results in a way that exhibits undesirable or even wrongful forms of bias (\citeauthor{noble2018algorithms}). 
Accordingly, mitigation of biases present in recommender systems is an area of active research. In the literature, this problem has been addressed either by debiasing a word embedding (\citeauthor{bolukbasi2016man}, \citeauthor{kaneko2019gender}, \citeauthor{stanczak2021survey}), debiasing contextual embedding (\citeauthor{bartl-etal-2020-unmasking}, \citeauthor{bartl2020unmasking}) or by re-ranking search results to eliminate such bias (\citeauthor{zehlike2020reducing}, \citeauthor{zehlike2020fairsearch},\citeauthor{raj2020comparing}). 

While the existing approaches of debiasing search terms or re-ranking search results are important approaches, there are times when forcing such a debiasing or re-ranking on the user may be undesirable.  For example, in politically charged queries, it is quite possible that a user \textit{wants} results that disproportionately represent one political party; similarly, when searching for information on vaccines, one may not wish to have balance between pro- and anti-vaccine views.  In such cases, providing a query recommendation is a `gentler' alternative to a behind-the-scenes debiasing, because it allows the user to decide whether she wants to see different results. 

In earlier work, we presented the \texttt{BalancedQR} framework \cite{10.1007/978-3-031-43421-1_25}, which addresses the alternative problem of balanced query recommendation. In this problem, an algorithm suggests less biased or oppositely biased alternatives to a query. \texttt{BalancedQR} uses a word embedding and an LLM to identify related queries, returning a Pareto front of high relevance, low bias recommendations. 

In that work, to evaluate \texttt{BalancedQR}, we applied it to scraped Reddit and Twitter datasets and considered bias along one axis (based on the sources of the tweet or the subreddit in which the document was posted). While that analysis was a starting point for  evaluating \texttt{BalancedQR}, it suffered from some important limitations.  Most importantly, bias was evaluated along only one dimension, while in many real-world search applications, documents can exhibit multiple forms of bias.  

Our goal here is to demonstrate use of \texttt{BalancedQR} on a dataset that is more closely aligned with what one might expect in a real-world search application.  Here, we apply the \texttt{BalancedQR} framework on a collection of Wikipedia documents provided by the 2022 TREC Fair Ranking Track. Along with the text of the documents, the dataset provides multiple dimensions of protected attributes (including geography, gender, and so on), as well as a set of related queries and relevant document judgments for each query. 
Additionally, Wikipedia is an important search domain in itself, and so demonstrating use of a search-related algorithm on this dataset provides valuable insights.

The TREC Wikipedia dataset motivates two important extensions to our earlier work.  First, we discuss how to extend \texttt{BalancedQR} to handle multiple dimensions of bias, such as gender, ethnicity, and geography. Second, while our earlier work only considered one method for generating candidate queries, here we consider multiple ways to generate candidate queries, and propose an evaluation framework to compare these methods. 


This paper is structured as follows: In Section~\ref{sec:relwork}, we describe related work.  In Section~\ref{sec:background}, we provide a description of \texttt{BalancedQR} and necessary extensions.  In Section~\ref{sec:dataset}, we give an overview of the TREC Wikipedia dataset.  In Sections~\ref{sec:setup} and~\ref{sec:results}, we describe the experimental setup and results.  We end with concluding thoughts.

\section{Related Work}\label{sec:relwork}
In this paper, we describe work in areas such as query expansion recommendation and fairness. 

\subsection{Query Expansion}
At the application level, the \texttt{BalancedQR} framework may seem similar to query expansion methods, as both produce a new query recommendation for the user. However, these are fundamentally different approaches. Query expansion adds new terms to the query based on either global methods, like the use of a thesaurus to find similar words for words in the query, or a local method like creating an automatic thesaurus based on a collection of documents. Query expansion on its own is effective in increasing recall, and with the use of relevance judgments on documents retrieved for the query, it can also help with increased precision ~\cite{christopher2008introduction}. In contrast, the \texttt{BalancedQR} framework is intended to show a diverse set of results to the user by recommending queries that can deliver different viewpoints on a topic.

Query expansion often relies on relevance feedback on retrieved documents for better results. Because asking users to provide relevance feedback on each retrieved document is a slow and time-consuming task, pseudo-relevance feedback is used, where a subset of top-k documents is deemed relevant and they are used to create a new expanded query (\citeauthor{mitra1998improving,10.1145/2071389.2071390,10.1145/2911451.2911539}). Using pseudo-relevance feedback to generate a new query also leads to results that are similar to the top results for the original query. This process is different from \texttt{BalancedQR}, where the framework may deliver queries that are only broadly related to the topics of initial search results retrieved for the user's query. 

Large language models with different prompting strategies have also been proposed for query expansion \citeauthor{jagerman2023query}. Zero-shot, few-shot, and Chain-of-Thought prompting strategies, along with pseudo-relevant documents, are used to generate new query terms. They also rely on a set of retrieved documents to find new terms, and therefore, they too suffer from similar drawbacks that are discussed above.

The traditional similarity-based approaches of query expansion and query rewriting methods are different than what we are trying to achieve with \texttt{BalancedQR} framework, where recommended queries can deliver a very diverse set of results to the user. For example, using query expansion methods on Wikipedia data for a query and relevance judgments if available, will lead to expanded queries that try to retrieve more relevant documents. Whereas, \texttt{BalancedQR} will recommend queries that can highlight completely different results that were not considered relevant for the original query. As we will see in Section \ref{sec:results}, querying `Politics' on the Wikipedia data leads to only 1 page out of 20 with a gender label associated with it. On the other hand, \texttt{BalancedQR} recommends multiple queries that lead to a significantly higher ratio of pages with gender labels.

\subsection{Fairness}
Existing methods look to debias the results either by debiasing word embeddings (\citeauthor{bolukbasi2016man}, \citeauthor{kaneko2019gender}) or to re-rank the results based on protected attributes (\citeauthor{Meike}) addresses group fairness concerns in rankings with greedy algorithms. (\citeauthor{liu2018personalizing}) presented re-ranking algorithms that balance personalization with fairness. (\citeauthor{touileb-etal-2021-using}) added metadata information, such as gender information, to the dataset to mitigate biases in classification models. (\citeauthor{10.1145/3582768.3582804}) used a variational autoencoder to debias word embeddings by using the learned latent space of the embedding. The \texttt{BalancedQR} framework differentiates itself from existing fairness-focused methods in that it neither alters ranking algorithms nor modifies the vector representations of data. \texttt{BalancedQR} framework works within a given search engine pipeline and delivers entirely new sets of queries with almost no modification to the existing architecture.

\section{Overview of \texttt{BalancedQR} and Extensions} 
\label{sec:background}

Here, we provide a brief overview of the balanced query recommendation framework, \texttt{BalancedQR}~ \cite{10.1007/978-3-031-43421-1_25}, which provides a set of query recommendations to the end user that are \textit{relevant} to the user's original search query, and exhibit greater \textit{diversity}.  For example, if the original query produced results with a strong male bias, the alternatives should be less so, or even exhibit a female bias. 

\subsection{Original \texttt{BalancedQR}}
\subsubsection{Problem Setup}
Search engine may return results to a user's query that are biased with respect to some attribute. These attributes can be those traditionally considered `protected', such as gender or race. However, bias can also appear for other attributes of interest, such as person's political alignment or personal interests etc. 

Formally, the goal of balanced query recommendation is to take the user's original query and recommend a set of \textit{balanced} queries. Each recommended query exhibits greater diversity and provides results that are relevant to the results returned for the user's original query. The `diversity' is based on selected attributes of the documents. We provide more discussions on `diversity' and `relevance' below.

Similar to fair re-ranking methods that aim to curate a less biased set of results, balanced query recommendation also aims to deliver more diverse, less biased results to the end user. But, unlike other direct debiasing methods, the \texttt{BalancedQR} results are not imposed on the user, and the user may choose whether to accept the \texttt{BalancedQR} recommended query. 

\subsubsection{\texttt{BalancedQR} Framework}

At a high level, \texttt{BalancedQR} performs the following steps:

(1) \texttt{BalancedQR} fetches the top search results for a query using an existing search engine algorithm.

(2) Then, \texttt{BalancedQR} identifies a set of alternative keywords nearest to words in the query using a word embedding model. It then prompts a large language model with the original query and alternative keywords to generate a set of candidate queries. \texttt{BalancedQR} then computes the bias and relevance scores for each query, where relevance is measured with respect to the original query (details below). Using the bias and relevance scores for each query, a Pareto front is calculated. The Pareto front contains a query if it is not dominated by any other query in the set. A query is non-dominated if there is no other candidate query that has higher relevance and a lower bias score. In some cases, we may use a `pseudo'-Pareto front that also considers queries that have higher bias but in the opposite direction.

(3) The above step is repeated by \texttt{BalancedQR} until a satisfactory Pareto front has been defined. Queries on the Pareto front are then recommended to the user.

\textbf{Measuring Diversity:} \texttt{BalancedQR} measures diversity in terms of bias. Documents were labeled with one value for bias based on either the subreddit it was posted on or the source of the news article, according to \url{www.allsides.com}. Each document returned from the search engine for a query has a bias value of +1 or -1 assigned to it. The bias for the query is then simply the average of biases of the returned documents.

\textbf{Measuring Relevance:} \label{subsec:relevance} We define the \textit{relevance} of a candidate query to an original query in terms of the similarity between the document sets returned for each query. \texttt{BalancedQR} uses Cosine similarity (bag of words representation), in which for each document in the two sets, we find the most similar document in the other set, and so define a mean Cosine similarity for each set. The overall similarity is the harmonic mean of these values (similar to a modified F1-score). 

\textbf{Limitations:} Our earlier work had a few important limitations: It was limited to supporting bias computation only along one axis. In many cases, a document is biased along multiple dimensions. It can be biased against gender, race, nationality, or other protected attributes. The second limitation was the absence of an evaluation framework to compare different query generation methods, such as word embedding, zero-shot LLMs, etc. 

We address these limitations in this work using the TREC Wikipedia dataset. The dataset contains useful protected attributes information for each document in the corpus as well as relevance judgments, quality score that helps with extending \texttt{BalancedQR} to handle biased results along multiple dimensions. The dataset also provides a set of queries that are used in this work to evaluate different query generation methods.

\subsection{Extensions}
\label{sec:pareto front}
\texttt{BalancedQR} \cite{10.1007/978-3-031-43421-1_25} suffered from fundamental algorithmic challenges that we overcome in this work by proposing a solution to handle biases along multiple dimensions as well as setting up an evaluation framework to test different query generation methods.

As discussed in Section~\ref{sec:background}, \texttt{BalancedQR} only handled bias along one dimension. But as we know, a query (by way of results returned for the query) can be biased along multiple attributes such as gender, race, nationality, etc. In this work, we score each query along the selected biased attribute dimensions. For example, if for a query and a dataset, we decide there are 5 biased (and/or protected) attributes, then we measure 5 attribute bias scores.
\begin{equation}
     Bias(query)= [biasScore_{dim1}, biasScore_{dim2}, \dots, biasScore_{dim5}]
\end{equation}
Here, each $biasScore$ is measured by the attribute labels of results for the query.  For example. if a total of ten results are retrieved for query $q$, out of which two have a political bias of $+1$ and the rest have a bias of $-1$, then the score for $q$ along the `political' attribute dimension is $-8$.  

Using the approach same as \cite{10.1007/978-3-031-43421-1_25} and describe here in section \ref{sec:background}, we also score each query on a relevance dimension and measure a relevance score $Rel(query)$. 

We now present notations in Table~\ref{tab:inputs} used in the \texttt{BalancedQR} framework, followed by the \texttt{BalancedQR} algorithm, shown in Algorithm~\ref{alg:BQR}, and the modified Pareto Front Algorithm, shown in Algorithm~\ref{alg:optimal_candidates}, to handle multiple biases. 

\begin{table}
\caption{Collective Inputs and Outputs of Algorithm}
\label{tab:inputs}
\centering
\begin{tabular}{c|l}
\hline
\multirow{2}{*}{Inputs} & \textbf{\emph{$Q$}}: Input Query \\
& \textbf{\emph{$Q_i$}}: Alternative Query\\
& \textbf{\emph{$d, D$}}: Document, set of documents \\
& \textbf{\emph{$S$}}: Search algorithm \\
& \textbf{\emph{$S(Q)$}}: Top-$n$ most relevant documents to query $Q$\\
&\hspace{1.25cm}from document set $D$, as found by algorithm $S$\\
& \textbf{\emph{$g(d), g(D)$}}: Diversity scores lists of a document $d$ \\
&\hspace{1.25cm}or document set $D$\\
& \textbf{\emph{$Rel(d), Rel(D)$}}: Relevance of a document $d$ or document\\
& \hspace{1.8cm} set $D$ to query $Q$\\
&\textbf{\emph{$W$}}: Word embedding \\
Outputs & \textbf{\emph{recs}}: final output set of candidate queries\\
\hline
\end{tabular}
\end{table}

\begin{algorithm}[t]

	\caption{Balanced query recommendation} 
    \label{alg:BQR}

	\begin{algorithmic}[1]
	\State $Q$ = original query
	\State $k$ = number of desired queries, $n$ = number of returned documents
	\State $max\_iter$ = maximum number of iterations, $num\_iters = 0$
	\State $Bias_Q = g_Q(D)$
	\State $sim$ = list of $k$ associated/ related words
    \If{$Q$ is multi-word query} 
        \State $sim$ = list of LLM({\em $w'$}) for each {\em $w'$} in sim
    \EndIf
	\State $recs = \{Q\}$
	\While {$|num\_iters| < max\_iter$ and $|recs|<k$}
    \For {each query {\em w'} in $sim$}
        \State $S_{w'}(D, n)$ = top-$n$ relevant documents from $D$ for {\em $w'$}
	    \State $Bias_{w'} = g_{w'}(D)$
	    \State $Rel(w')$ = F1-score between $S_{w'}(D, n)$ and $S_{Q}(D, n)$
	    \If {$w'$ is not dominated by any query in $recs$} 
	    \State Add $w'$ to {$recs$}
	    \State Remove queries from $recs$ that are dominated by $w'$
	    \EndIf
	\EndFor
	\State $sim$ = \{next most similar word from word embedding\}
	\State $num\_iters ++$
	\EndWhile
    \State Return $recs$
	\end{algorithmic} 
\end{algorithm}

\begin{algorithm}[t]
\caption{Pareto Front algorithm to Find Balanced Queries}
\label{alg:optimal_candidates}
\begin{algorithmic}[1]
\State $res$ = set of candidate queries
\State $Q$ = original query
\state $k$ = number of biased attributes
\State $ans$ = $\{\}$

\For{each index $i$, query $q_i'$ in $res$} 
    \State $isDominant$ =  False
    \State $Bias(q_i')$ = $[Bias_1{q_i'}(D), Bias_2{q_i'}(D)\dots, Bias_k{q_i'}(D)]$
    
    \State $Rel(q_i')$ =  F1-score between $S_{q'}(D, n)$ and $S_{Q}(D, n)$
    \State $dimScores(q_i')$ = $Bias(q_i')$ + $[Rel(q_i')]$
    \For{each index $j$, query $q_j'$ in $res$}
        \If{index $i$ != index $j$ and 
           \\ $dimScores(q_j').all() \geq dimScores(q_i')$ and \\
            $dimScores(q_j').any() > dimScores(q_i')$}
            \State $isDominant$ =  True
            \State \textbf{break}  
        \EndIf
    \EndFor
    \If{$isDominant$ == False}
        \State Add $q_i'$ to $ans$ 
    \EndIf
\EndFor
\State Return $ans$
\end{algorithmic}
\end{algorithm}

\section{Dataset}
\label{sec:dataset}
In this paper, we use a training set of Wikipedia documents made available by the 2022 TREC Fair Ranking Track  \cite{ekstrand2021trec}. We use the dataset for the analysis as it contains rich information about attributes that can used for \texttt{BalancedQR} framework, it also provides meta data information that can be used for future algorithmic innovations along the lines of new evaluation metrics, personalization of results and to verify precision/ recall of new methods.
Along with the text in the dataset, the corpus also contains attribute information such as Geographic location (article topic), Gender (biographies only), Quality Score, etc. The track also provides a collection of queries (46 in total for the training set), where each query row contains the title of the query topic, associated keywords, and a list of documents relevant to that topic. We provide attribute information available for documents and queries in table \ref{ref:dataTable}. We use \textit{Geographic location (article topic), Gender (biographies only), and relevance} as attribute dimensions for \texttt{BalancedQR}. We refer to the TREC page \cite{ekstrand2021trec} for more information on the dataset.  

\begin{table}
    \centering
    \begin{tabular}{cl}
        Collection & Attributes\\
        Corpus & id, title, url, text, Geographic location (article topic), \\
               & Geographic location (article sources), \\
               & Gender (biographies), Age of the topic, \\
               & Occupation (biographies), Alphabetical, Age of the article \\
               & Popularity (pageviews), Replication of articles \\
        Queries & id,  title, keywords, relevant docs\\ 
    \end{tabular}
    \caption{Attributes information made available for dataset used in TREC Fair Ranking Track }
    \label{ref:dataTable}
\end{table}

\section{Experimental Setup}\label{sec:setup}
We implement \texttt{BalancedQR} framework using the setup mention in Section \ref{sec:BQRdetails}. For a set of query topics and candidate queries generated from different query generation methods, we retrieve a set of balanced queries to recommend.
\subsection{Candidate Query Evaluation Framework}
\label{ref:Candidate Query Evaluation Framework}
One of the most crucial parts of \texttt{BalancedQR} is to find a good set $sim$ of candidate queries that can be filtered out later in the process to get a set $recs$ of balanced queries that can be recommended to the user.

There can be multiple ways to generate the candidate queryset $sim$. We implement three methods to generate queries for the \texttt{BalancedQR} framework.

Method 1: We use GloVe embedding \cite{pennington2014glove} developed from wikipedia dataset. For a given query topic, we retrieve the ten most similar words using GloVe wiki word embeddings, and these similar words are then treated as our candidate queries.

Method 2: We use the most similar words fetched in step (1) and prompt a large language model to generate a list of alternative queries. A snippet of the prompt used is \textit{Please generate 10 different search queries by only using Topic and associated keywords with the following guidelines. Make sure the queries don't have keywords other than what's provided here}. We use the most similar words fetched in Method 1 as our associated keywords.

Method 3: For a given query topic, we leverage the list of associated keywords available in the dataset (as detailed in Section \ref{sec:dataset}). We prompt a Large Language Model (LLM) to utilize both these associated keywords and the query topic to generate a list of candidate queries.
 We leverage the Pareto front approach and propose the following steps for the evaluation of candidate query generation methods. 
\begin{enumerate}
    \item Let $recs_i$ be the recommended queries from the \texttt{BalancedQR} framework, where the candidate queries were originally generated by method $m_i$.
    
    \item Each query $q \in recs_i$ has the relevance and bias scores calculated for each of the attribute dimensions. 
    
    $dimScores(q)$ = $[Bias_1(q), Bias_2(q), \dots,Bias_k(q), Rel(q)]$

    \item To compare $recs_1$ (from method $m_1$) against $recs_2$ (from method $m_2$), we evaluate how many queries in $recs_1$ are dominated by queries in $recs_2$. Specifically, for each query $q_2 \in recs_2$, we count the number of queries $q_1 \in recs_1$  such that $q_2$ dominates $q_1$ based on their respective $dimScores$. We then sum these counts to get a total domination score for $recs2$ and $m_2$ over $recs1$ and $m_1$. (To reiterate, use of dominance here refers to total number of queries generated from one method dominated on the Pareto front by each query generated from other method.)
\end{enumerate}

\subsection{\texttt{BalancedQR Implementation}}
\label{sec:BQRdetails} 
Using search results for the candidate query and the results for the original query, we calculate a modified F1-score to assign the $relevance$ score to each candidate query. We explain $relevance$ measure in detail in Section \ref{sec:background}. 

We use Jensen-Shannon divergence \cite{lin2002divergence} to separately measure the difference in geographic and gender distributions in results returned for the original query and results from the candidate query. Jensen-Shannon divergence is bounded between $0\leq JSD(C||Q) \leq 1$, where $C, Q$ are two probability distributions. Here, 0 means $C$ and $Q$ are identical distributions and 1 means the two probability distributions are entirely dissimilar.


We assign a \textit{geo entropy} score to a candidate query based on the difference in geographic distributions between results for the candidate query and results from the original query. Similarly, we assign a \textit{gender entropy} score to the candidate query by comparing gender distributions in candidate query results with results from the original query. 

Each original query and its candidate query are searched against the wiki corpus using the Pyserini framework implementation of BM25 \cite{Lin_etal_SIGIR2021_Pyserini} algorithm. We process the results such that each query will have dimension scores, $dimScores$ of the form:

$[geoEntropy, genderEntropy, relevance]$.

We then use \texttt{BalancedQR} framework with the modified Pareto front model defined in Section \ref{sec:pareto front} with $relevance$, \textit{geo entropy}, and \textit{gender entropy} as the attribute dimensions to recommend a set of balanced queries to the user. For example, If a query returns Wikipedia pages that are heavily biased towards one geographic region such as North America then \texttt{BalancedQR} can recommend queries that will direct the user to view relevant pages from other geographic regions if the user decides to accept the recommended \texttt{BalancedQR} query.

\section{Results}
\label{sec:results}
We first evaluate the various candidate query generation methods described earlier. Recall that \textbf{Method 1} directly uses the nearest words from the word embedding model as candidate queries. For example, candidate queries corresponding to an original query `classical music' are `jazz' and `concert'.  \textbf{Method 2} uses the query and the nearest words from the word embedding model to prompt the LLM to generate candidate queries. For example, For original query `classical music', we get candidate queries like `Classical music compositions', `Classical music concert recording'.  Lastly,  \textbf{Method 3} uses the original query and the associated keywords available in the dataset and, like Method 2, prompts the LLM to generate candidate queries. For original query `classical music' using Method 3, we generate candidate queries such as `bach violin cantata' and `haydn symphony orchestra'. We note that Method 3 is only possible because the dataset provides associated keywords for each query. 
\begin{figure}
    \centering
    \includegraphics[width=1.0\linewidth]{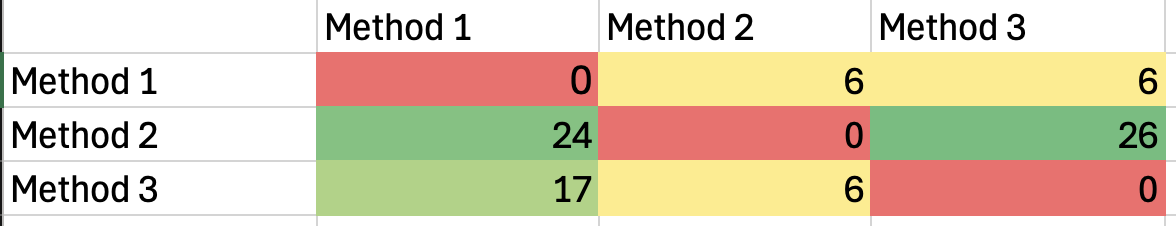}
    \caption{Dominance results for each method on 10 randomly selected query topics. We see that Method 2 dominates the other two methods. The high dominance scores are highlighted in green and lowest are in red. Scores in-between low and high are highlighted with yellow and lighter shades of green.}
    \label{fig:methodEvalResults}
\end{figure}

Figure~\ref{fig:methodEvalResults} shows that Method 2, which involves prompting a large language model with the original query and similar words retrieved from GloVe wiki word embeddings to generate candidate queries, generally outperforms the other two methods. We also observe that in most of the examples, Method 1, which directly uses a list of similar words from the word embedding to identify candidate queries, rarely dominates the other two methods. 

We now want to highlight why the \texttt{BalancedQR} queries are relevant when using a corpus like Wikipedia. Wikipedia, a vast platform built and maintained by volunteers, often reflects the historical biases in internet access in terms of volunteers, editors or usage. Europe or North America can account for more than $80$ percent of geo tagged articles on Wikipedia \footnote{\url{https://en.wikipedia.org/wiki/Geographical_bias_on_Wikipedia}}. Consider a user from South Asia or Middle Africa who is interested in learning more about \textit{politics}. When she searches the term, nearly all returned results are from regions like Northern America, Europe as most of mainstream \textit{famous} political pages are from the west and they are viewed much more by people again, from the same part of the world. The content may be abstractly relevant to her need but \texttt{BalancedQR} can do better by providing alternative queries that can lead her to information more relevant to their specific regions, along with the original mainstream results. Ensuring a more diverse and representative search experience.
We only discuss geographic bias on Wikipedia and relevance of \texttt{BalancedQR} above. But similar arguments can also be made for gender bias issues on the platform \footnote{\url{https://www.newstatesman.com/long-reads/2015/05/wikipedia-has-colossal-problem-women-dont-edit-it}}.

\begin{figure}
    \centering
    \includegraphics[width=1.0\linewidth]{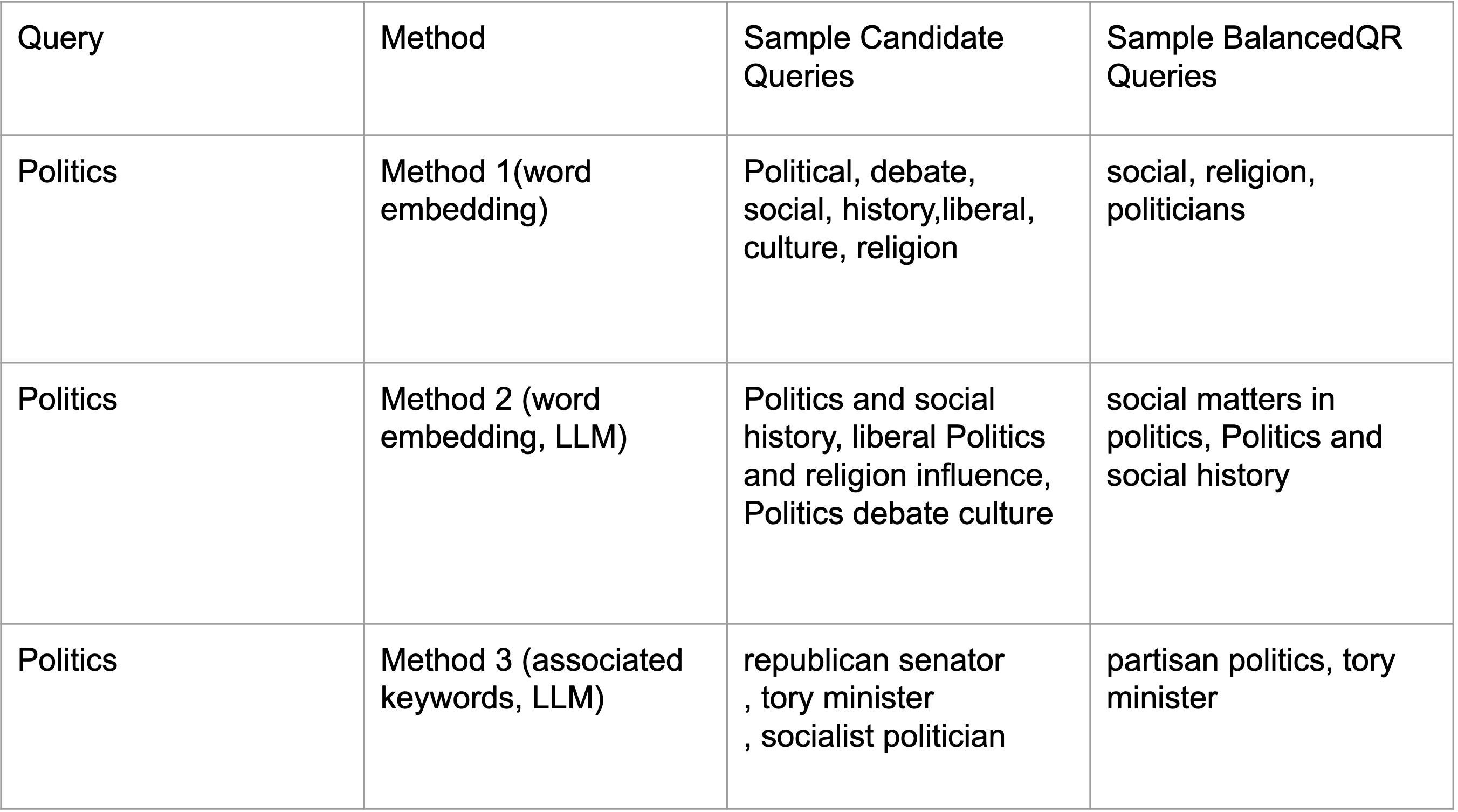}
    \caption{Sample candidate and \texttt{BalancedQR}-recommended queries using word embeddings (Method 1), word embedding swith a LLM (Method 2) and using given keywords from the dataset with LLM to generate queries (Method 3). }
    \label{fig:politics}
\end{figure}

\begin{figure}
    \centering
    \includegraphics[width=1.0\linewidth]{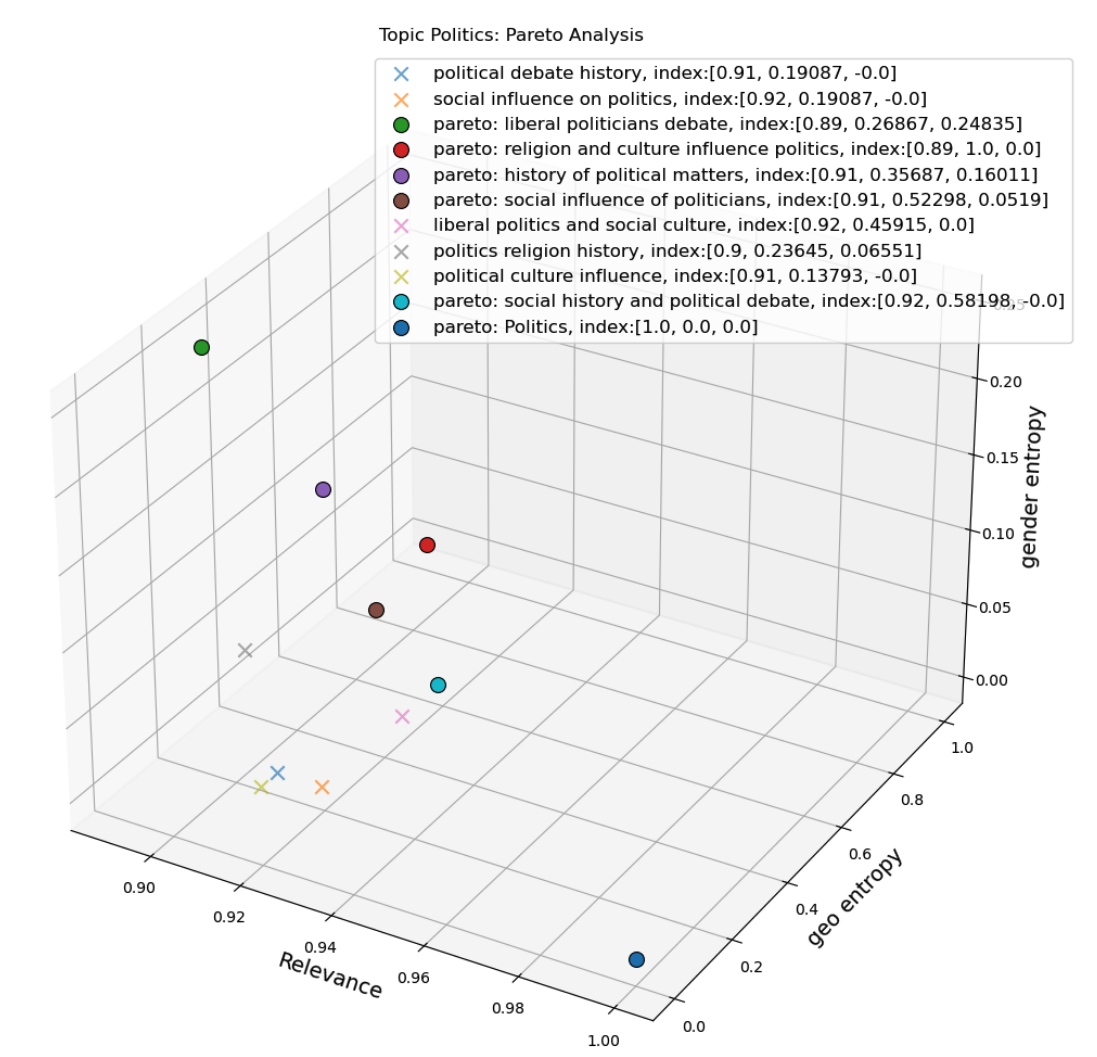}
    \caption{\texttt{BalancedQR} applied on query `politics' using  most similar words from wiki GloVe word embeddings and Gemini LLM to generate candidate queries. Queries recommended by \texttt{BalancedQR} are highlighted in legend as `pareto' and queries not recommended are marked as 'x' in the legend and plot. The values besides the queries in legend are relevance, geographic entropy and gender entropy scores.}
    \label{fig:m2-politics}
\end{figure}

Next, we provide qualitative analysis on query `Politics' using Method 2 in \texttt{BalancedQR} framework in Figures~[\ref{fig:politics}, \ref{fig:m2-politics}]


We see in Figure~\ref{fig:m2-politics} that the original query `Politics' has a relevance of 1 and no entropy scores. This is expected, as the dimension scores for the original query are derived by comparing the same set of results against itself. For this query, \texttt{BalancedQR} generates candidate queries such as `social influence on politics', `political culture influence', and `social history and political debate', using the GloVe wiki word embedding and the Gemini large language model. 

We observe that most candidate queries have high geographic entropy scores, meaning that the returned result set of documents refers to multiple geographic regions. Analysis of the resulting document set reveals that the results for the original query referenced the `Northern America' region twice out of a total twenty documents, rest of the returned documents did not have a geography label associated with them. The first document that referenced the `Northern America' region was a biography page of `David Easton' a Canadian-born American political scientist and the second document was a list of advocacy groups in Canada concerning political social interests.   In contrast, the candidate query `religion and culture influence politics' has a \textit{geo entropy} score of 1 which shows that the candidate query geographic distribution was entirely dissimilar to the geographic distribution of the original query. The returned documents for the candidate query reference pages such as, a page about influence of ethnic, religious, multicultural background of Indonesia on its politics, another page about diverse culture of Cameroon and norms of Cameroon politics. Again, from the result analysis, we learn that the returned result set is still relevant to \textit{politics} and now the query references multiple regions, such as `Eastern Asia' thrice, `Southern Asia' twice, and other regions like `Northern Europe' and `South-eastern Asia' once. 

Note that many candidate queries refer to `social', `culture', `debate', and `religion'. 
We hypothesize that, in this case, results for `Politics' do not contain references to \textit{religion} or \textit{culture},and when the query is specific to regions that have high involvement of \textit{religion} or \textit{culture}, then the returned results are more geographically diverse.

Similarly, we find from document analysis that the original query `Politics' results only refer to the `male' gender once, the `David Easton' biography page that we mentioned above.  However, certain candidate queries show greater amounts of gender entropy. For example, the query `liberal politicians debate' has a gender entropy score of 0.248, and we confirm through examination of the results that the document set contains results that refer to the `male' gender seven times and the `female' gender five times. It returns biography pages for male politicans such as `Kevin Clark' from Canada and Swedish politician `Carl Gustaf Ekman'. Retrieved pages also reference female politicians like `Alexandra Mendè' from Canada, `Bronwyn Bishop' and `Karina Okotel' from Australia and New Zealand. We know that the gender attribute is only present for biographies in the dataset, so it is possible that the candidate query `liberal politicians debate' is able to retrieve gender diverse biography pages in the result set, in comparison to the original query `Politics', which returned results containing only one male biography page.

We observe that the \texttt{BalancedQR} framework returns queries such as `liberal politicians debate', `social influence on politicians', and `history of political matters' as balanced queries to the user, using which the user can view more diverse results than what are retrieved with the original query `Politics'.

We also highlight how the subtle choice of words in the query can lead to a more diverse set of results. Using words such as `religion', `culture' in the context of politics can lead to more geographically diverse results, referring to places where religion or cultural norms play or have played a major part in politics. For the gender dimension, results are comparatively unclear, but we see that more gender diverse results are retrieved for queries that use `politicians', `political' in the query, compared to the word `politics'. This may be because when we query `politicians' or  `political', then more biographies are fetched as we refer to political personalities, as compared to when we have a query that refers to the field of `politics' abstractly.


\section{Conclusion and Future Work}
In future work, we aim to deliver personalized results using the \texttt{BalancedQR} framework. This is an important limitation that we want to rectify in the future. Some users may be interested in receiving queries that can deliver a diverse set of results, while some users may not want such results. We want to modify \texttt{BalancedQR} framework such that it can respect users interests. \texttt{BalancedQR} can leverage multi-arm bandit approaches such as the UCB1 algorithm to learn user preferences based on user history and deliver a more personalized set of queries to the user. 

Other query recommendation methods, such as query expansion and query reformulation, are based on precision measures that aim to deliver better but similar types of results to the user. Unlike them, \texttt{BalancedQR} framework cannot be evaluated comprehensively in a lab setting as it tries to deliver queries that may be relevant to the user's original query, but it will also provide results from diverse points of views, different than what the user intended with the original query. A user survey can be conducted to conclude if \texttt{BalancedQR} queries are ultimately useful to the user. 

A limitation of the \texttt{BalancedQR} framework currently, is the reliance on LLMs to generate candidate queries. LLMs can be fickle and subject to biases from external sources not in control of the dataset used in the paper or the fairness algorithms, We aim to reduce the use of LLMs in \texttt{BalancedQR} in our future work.

In this paper, We extended \texttt{BalancedQR} framework to deal with the challenges of multiple biased attributes present in Wikipedia documents. We also established a evaluation framework to judge different candidate query generation methods. Along with results re-ranking and debiasing word embeddings, we show that balanced query recommendation is also a way to amplify diverse and relevant results to the user. \texttt{BalancedQR} recommended queries can help user see multiple sides of a topic. A major application of this framework is in mitigating the harmful effects of social media, where balanced queries can help in fighting the rising tensions due to filter bubbles and echo chambers online.

\bibliographystyle{ACM-Reference-Format}
\bibliography{draft}


\begin{thebibliography}{23}


\ifx \showCODEN    \undefined \def \showCODEN     #1{\unskip}     \fi
\ifx \showISBNx    \undefined \def \showISBNx     #1{\unskip}     \fi
\ifx \showISBNxiii \undefined \def \showISBNxiii  #1{\unskip}     \fi
\ifx \showISSN     \undefined \def \showISSN      #1{\unskip}     \fi
\ifx \showLCCN     \undefined \def \showLCCN      #1{\unskip}     \fi
\ifx \shownote     \undefined \def \shownote      #1{#1}          \fi
\ifx \showarticletitle \undefined \def \showarticletitle #1{#1}   \fi
\ifx \showURL      \undefined \def \showURL       {\relax}        \fi
\providecommand\bibfield[2]{#2}
\providecommand\bibinfo[2]{#2}
\providecommand\natexlab[1]{#1}
\providecommand\showeprint[2][]{arXiv:#2}

\bibitem[Bartl et~al\mbox{.}(2020a)]%
        {bartl-etal-2020-unmasking}
\bibfield{author}{\bibinfo{person}{Marion Bartl}, \bibinfo{person}{Malvina Nissim}, {and} \bibinfo{person}{Albert Gatt}.} \bibinfo{year}{2020}\natexlab{a}.
\newblock \showarticletitle{Unmasking Contextual Stereotypes: Measuring and Mitigating {BERT}{'}s Gender Bias}. In \bibinfo{booktitle}{\emph{Proceedings of the Second Workshop on Gender Bias in Natural Language Processing}}, \bibfield{editor}{\bibinfo{person}{Marta~R. Costa-juss{\`a}}, \bibinfo{person}{Christian Hardmeier}, \bibinfo{person}{Will Radford}, {and} \bibinfo{person}{Kellie Webster}} (Eds.). \bibinfo{publisher}{Association for Computational Linguistics}, \bibinfo{address}{Barcelona, Spain (Online)}, \bibinfo{pages}{1--16}.
\newblock
\urldef\tempurl%
\url{https://aclanthology.org/2020.gebnlp-1.1}
\showURL{%
\tempurl}


\bibitem[Bartl et~al\mbox{.}(2020b)]%
        {bartl2020unmasking}
\bibfield{author}{\bibinfo{person}{Marion Bartl}, \bibinfo{person}{Malvina Nissim}, {and} \bibinfo{person}{Albert Gatt}.} \bibinfo{year}{2020}\natexlab{b}.
\newblock \showarticletitle{Unmasking contextual stereotypes: Measuring and mitigating BERT's gender bias}.
\newblock \bibinfo{journal}{\emph{arXiv preprint arXiv:2010.14534}} (\bibinfo{year}{2020}).
\newblock


\bibitem[Bolukbasi et~al\mbox{.}(2016)]%
        {bolukbasi2016man}
\bibfield{author}{\bibinfo{person}{Tolga Bolukbasi}, \bibinfo{person}{Kai-Wei Chang}, \bibinfo{person}{James~Y Zou}, \bibinfo{person}{Venkatesh Saligrama}, {and} \bibinfo{person}{Adam~T Kalai}.} \bibinfo{year}{2016}\natexlab{}.
\newblock \showarticletitle{Man is to computer programmer as woman is to homemaker? debiasing word embeddings}.
\newblock \bibinfo{journal}{\emph{Advances in neural information processing systems}}  \bibinfo{volume}{29} (\bibinfo{year}{2016}).
\newblock


\bibitem[Carpineto and Romano(2012)]%
        {10.1145/2071389.2071390}
\bibfield{author}{\bibinfo{person}{Claudio Carpineto} {and} \bibinfo{person}{Giovanni Romano}.} \bibinfo{year}{2012}\natexlab{}.
\newblock \showarticletitle{A Survey of Automatic Query Expansion in Information Retrieval}.
\newblock \bibinfo{journal}{\emph{ACM Comput. Surv.}} \bibinfo{volume}{44}, \bibinfo{number}{1}, Article \bibinfo{articleno}{1} (\bibinfo{date}{jan} \bibinfo{year}{2012}), \bibinfo{numpages}{50}~pages.
\newblock
\showISSN{0360-0300}
\href{https://doi.org/10.1145/2071389.2071390}{doi:\nolinkurl{10.1145/2071389.2071390}}


\bibitem[Christopher et~al\mbox{.}(2008)]%
        {christopher2008introduction}
\bibfield{author}{\bibinfo{person}{D~Manning Christopher}, \bibinfo{person}{Raghavan Prabhakar}, {and} \bibinfo{person}{Schutze Hinrich}.} \bibinfo{year}{2008}\natexlab{}.
\newblock \bibinfo{title}{Introduction to information retrieval}.
\newblock


\bibitem[Ekstrand et~al\mbox{.}(2021)]%
        {ekstrand2021trec}
\bibfield{author}{\bibinfo{person}{Michael~D Ekstrand}, \bibinfo{person}{Graham McDonald}, \bibinfo{person}{Amifa Raj}, \bibinfo{person}{Isaac Johnson}, {and} \bibinfo{person}{Morten Warncke-Wang}.} \bibinfo{year}{2021}\natexlab{}.
\newblock \bibinfo{title}{TREC 2022 fair ranking track participant instructions}.
\newblock


\bibitem[Jagerman et~al\mbox{.}(2023)]%
        {jagerman2023query}
\bibfield{author}{\bibinfo{person}{Rolf Jagerman}, \bibinfo{person}{Honglei Zhuang}, \bibinfo{person}{Zhen Qin}, \bibinfo{person}{Xuanhui Wang}, {and} \bibinfo{person}{Michael Bendersky}.} \bibinfo{year}{2023}\natexlab{}.
\newblock \showarticletitle{Query expansion by prompting large language models}.
\newblock \bibinfo{journal}{\emph{arXiv preprint arXiv:2305.03653}} (\bibinfo{year}{2023}).
\newblock


\bibitem[Kaneko and Bollegala(2019)]%
        {kaneko2019gender}
\bibfield{author}{\bibinfo{person}{Masahiro Kaneko} {and} \bibinfo{person}{Danushka Bollegala}.} \bibinfo{year}{2019}\natexlab{}.
\newblock \showarticletitle{Gender-preserving debiasing for pre-trained word embeddings}.
\newblock \bibinfo{journal}{\emph{arXiv preprint arXiv:1906.00742}} (\bibinfo{year}{2019}).
\newblock


\bibitem[Lin(2002)]%
        {lin2002divergence}
\bibfield{author}{\bibinfo{person}{Jianhua Lin}.} \bibinfo{year}{2002}\natexlab{}.
\newblock \showarticletitle{Divergence measures based on the Shannon entropy}.
\newblock \bibinfo{journal}{\emph{IEEE Transactions on Information theory}} \bibinfo{volume}{37}, \bibinfo{number}{1} (\bibinfo{year}{2002}), \bibinfo{pages}{145--151}.
\newblock


\bibitem[Lin et~al\mbox{.}(2021)]%
        {Lin_etal_SIGIR2021_Pyserini}
\bibfield{author}{\bibinfo{person}{Jimmy Lin}, \bibinfo{person}{Xueguang Ma}, \bibinfo{person}{Sheng-Chieh Lin}, \bibinfo{person}{Jheng-Hong Yang}, \bibinfo{person}{Ronak Pradeep}, {and} \bibinfo{person}{Rodrigo Nogueira}.} \bibinfo{year}{2021}\natexlab{}.
\newblock \showarticletitle{{Pyserini}: A {Python} Toolkit for Reproducible Information Retrieval Research with Sparse and Dense Representations}. In \bibinfo{booktitle}{\emph{Proceedings of the 44th Annual International ACM SIGIR Conference on Research and Development in Information Retrieval (SIGIR 2021)}}. \bibinfo{pages}{2356--2362}.
\newblock


\bibitem[Liu and Burke(2018)]%
        {liu2018personalizing}
\bibfield{author}{\bibinfo{person}{Weiwen Liu} {and} \bibinfo{person}{Robin Burke}.} \bibinfo{year}{2018}\natexlab{}.
\newblock \showarticletitle{Personalizing fairness-aware re-ranking}.
\newblock \bibinfo{journal}{\emph{arXiv preprint arXiv:1809.02921}} (\bibinfo{year}{2018}).
\newblock


\bibitem[Meike~Zehlike(2018)]%
        {Meike}
\bibfield{author}{\bibinfo{person}{Francesco Bonchi et~al. Meike~Zehlike}.} \bibinfo{year}{2018}\natexlab{}.
\newblock \showarticletitle{FA*IR: A Fair Top-k Ranking Algorithm}.
\newblock  (\bibinfo{year}{2018}).
\newblock
\newblock
\shownote{\url{https://arxiv.org/pdf/1706.06368.pdf}}.


\bibitem[Mishra and Soundarajan(2023)]%
        {10.1007/978-3-031-43421-1_25}
\bibfield{author}{\bibinfo{person}{Harshit Mishra} {and} \bibinfo{person}{Sucheta Soundarajan}.} \bibinfo{year}{2023}\natexlab{}.
\newblock \showarticletitle{BalancedQR: A Framework for Balanced Query Recommendation}. In \bibinfo{booktitle}{\emph{Machine Learning and Knowledge Discovery in Databases: Research Track}}, \bibfield{editor}{\bibinfo{person}{Danai Koutra}, \bibinfo{person}{Claudia Plant}, \bibinfo{person}{Manuel Gomez~Rodriguez}, \bibinfo{person}{Elena Baralis}, {and} \bibinfo{person}{Francesco Bonchi}} (Eds.). \bibinfo{publisher}{Springer Nature Switzerland}, \bibinfo{address}{Cham}, \bibinfo{pages}{420--435}.
\newblock
\showISBNx{978-3-031-43421-1}


\bibitem[Mitra et~al\mbox{.}(1998)]%
        {mitra1998improving}
\bibfield{author}{\bibinfo{person}{Mandar Mitra}, \bibinfo{person}{Amit Singhal}, {and} \bibinfo{person}{Chris Buckley}.} \bibinfo{year}{1998}\natexlab{}.
\newblock \showarticletitle{Improving automatic query expansion}. In \bibinfo{booktitle}{\emph{Proceedings of the 21st annual international ACM SIGIR conference on Research and development in information retrieval}}. \bibinfo{pages}{206--214}.
\newblock


\bibitem[Noble(2018)]%
        {noble2018algorithms}
\bibfield{author}{\bibinfo{person}{Safiya~Umoja Noble}.} \bibinfo{year}{2018}\natexlab{}.
\newblock \showarticletitle{Algorithms of oppression: How search engines reinforce racism}.
\newblock In \bibinfo{booktitle}{\emph{Algorithms of oppression}}. \bibinfo{publisher}{New York university press}.
\newblock


\bibitem[Pennington et~al\mbox{.}(2014)]%
        {pennington2014glove}
\bibfield{author}{\bibinfo{person}{Jeffrey Pennington}, \bibinfo{person}{Richard Socher}, {and} \bibinfo{person}{Christopher~D. Manning}.} \bibinfo{year}{2014}\natexlab{}.
\newblock \showarticletitle{GloVe: Global Vectors for Word Representation}. In \bibinfo{booktitle}{\emph{Empirical Methods in Natural Language Processing (EMNLP)}}. \bibinfo{pages}{1532--1543}.
\newblock
\urldef\tempurl%
\url{http://www.aclweb.org/anthology/D14-1162}
\showURL{%
\tempurl}


\bibitem[Raj and Ekstrand(2020)]%
        {raj2020comparing}
\bibfield{author}{\bibinfo{person}{Amifa Raj} {and} \bibinfo{person}{Michael~D Ekstrand}.} \bibinfo{year}{2020}\natexlab{}.
\newblock \showarticletitle{Comparing fair ranking metrics}.
\newblock \bibinfo{journal}{\emph{arXiv preprint arXiv:2009.01311}} (\bibinfo{year}{2020}).
\newblock


\bibitem[Stanczak and Augenstein(2021)]%
        {stanczak2021survey}
\bibfield{author}{\bibinfo{person}{Karolina Stanczak} {and} \bibinfo{person}{Isabelle Augenstein}.} \bibinfo{year}{2021}\natexlab{}.
\newblock \showarticletitle{A survey on gender bias in natural language processing}.
\newblock \bibinfo{journal}{\emph{arXiv preprint arXiv:2112.14168}} (\bibinfo{year}{2021}).
\newblock


\bibitem[Touileb et~al\mbox{.}(2021)]%
        {touileb-etal-2021-using}
\bibfield{author}{\bibinfo{person}{Samia Touileb}, \bibinfo{person}{Lilja {\O}vrelid}, {and} \bibinfo{person}{Erik Velldal}.} \bibinfo{year}{2021}\natexlab{}.
\newblock \showarticletitle{Using Gender- and Polarity-Informed Models to Investigate Bias}. In \bibinfo{booktitle}{\emph{Proceedings of the 3rd Workshop on Gender Bias in Natural Language Processing}}, \bibfield{editor}{\bibinfo{person}{Marta Costa-jussa}, \bibinfo{person}{Hila Gonen}, \bibinfo{person}{Christian Hardmeier}, {and} \bibinfo{person}{Kellie Webster}} (Eds.). \bibinfo{publisher}{Association for Computational Linguistics}, \bibinfo{address}{Online}, \bibinfo{pages}{66--74}.
\newblock
\href{https://doi.org/10.18653/v1/2021.gebnlp-1.8}{doi:\nolinkurl{10.18653/v1/2021.gebnlp-1.8}}


\bibitem[Tyagi et~al\mbox{.}(2023)]%
        {10.1145/3582768.3582804}
\bibfield{author}{\bibinfo{person}{Swati Tyagi}, \bibinfo{person}{Jiaheng Xie}, {and} \bibinfo{person}{Rick Andrews}.} \bibinfo{year}{2023}\natexlab{}.
\newblock \showarticletitle{E-VAN: Enhanced Variational AutoEncoder Network for Mitigating Gender Bias in Static Word Embeddings}. In \bibinfo{booktitle}{\emph{Proceedings of the 2022 6th International Conference on Natural Language Processing and Information Retrieval}} (<conf-loc>, <city>Bangkok</city>, <country>Thailand</country>, </conf-loc>) \emph{(\bibinfo{series}{NLPIR '22})}. \bibinfo{publisher}{Association for Computing Machinery}, \bibinfo{address}{New York, NY, USA}, \bibinfo{pages}{57–64}.
\newblock
\showISBNx{9781450397629}
\href{https://doi.org/10.1145/3582768.3582804}{doi:\nolinkurl{10.1145/3582768.3582804}}


\bibitem[Zehlike and Castillo(2020)]%
        {zehlike2020reducing}
\bibfield{author}{\bibinfo{person}{Meike Zehlike} {and} \bibinfo{person}{Carlos Castillo}.} \bibinfo{year}{2020}\natexlab{}.
\newblock \showarticletitle{Reducing disparate exposure in ranking: A learning to rank approach}. In \bibinfo{booktitle}{\emph{Proceedings of the web conference 2020}}. \bibinfo{pages}{2849--2855}.
\newblock


\bibitem[Zehlike et~al\mbox{.}(2020)]%
        {zehlike2020fairsearch}
\bibfield{author}{\bibinfo{person}{Meike Zehlike}, \bibinfo{person}{Tom S{\"u}hr}, \bibinfo{person}{Carlos Castillo}, {and} \bibinfo{person}{Ivan Kitanovski}.} \bibinfo{year}{2020}\natexlab{}.
\newblock \showarticletitle{Fairsearch: A tool for fairness in ranked search results}. In \bibinfo{booktitle}{\emph{Companion proceedings of the web conference 2020}}. \bibinfo{pages}{172--175}.
\newblock


\bibitem[Zhang et~al\mbox{.}(2016)]%
        {10.1145/2911451.2911539}
\bibfield{author}{\bibinfo{person}{Zhiwei Zhang}, \bibinfo{person}{Qifan Wang}, \bibinfo{person}{Luo Si}, {and} \bibinfo{person}{Jianfeng Gao}.} \bibinfo{year}{2016}\natexlab{}.
\newblock \showarticletitle{Learning for Efficient Supervised Query Expansion via Two-stage Feature Selection}. In \bibinfo{booktitle}{\emph{Proceedings of the 39th International ACM SIGIR Conference on Research and Development in Information Retrieval}} (<conf-loc>, <city>Pisa</city>, <country>Italy</country>, </conf-loc>) \emph{(\bibinfo{series}{SIGIR '16})}. \bibinfo{publisher}{Association for Computing Machinery}, \bibinfo{address}{New York, NY, USA}, \bibinfo{pages}{265–274}.
\newblock
\showISBNx{9781450340694}
\href{https://doi.org/10.1145/2911451.2911539}{doi:\nolinkurl{10.1145/2911451.2911539}}


\end{thebibliography}

\end{document}